\documentclass[prl ,twocolumn ,secnumarabic,amssymb, amsmath,nobibnotes,
showpacs,aps, prl]{revtex4}%
\usepackage{bm}
\usepackage{amsfonts}
\usepackage{amsmath}
\usepackage{amssymb}
\usepackage{graphicx}%
\expandafter\ifx\csname package@font\endcsname\relax\else
 \expandafter\expandafter
 \expandafter
\expandafter
 \expandafter{\csname package@font\endcsname}
\fi

\begin{document}

\title{ Stability under Perturbations of the Large Time Average  Motion of Dynamical Systems with Conserved Phase Space Volume. }

\author{Gy\"{o}rgy Steinbrecher}
\affiliation{Association EURATOM-MEdC, Physics Faculty, University of Craiova, Romania }

\author{Boris Weyssow}
\email{bweyssow@ulb.ac.be}
\affiliation{Association EURATOM-Etat Belge, Physique Statistique et Plasmas, Universit\'e
Libre de Bruxelles, Campus plaine, CP 231, B-1050 Bruxelles, Belgium.}


\begin{abstract}

The stability against perturbations of a dynamical system conserving a generalized phase-space volume is studied by exploiting the similarity between statistical physics formalism and that of ergodic theory. A general continuity theorem is proved. Resulting from this theorem the double average - time and ensemble - of an observable of a weakly perturbed ergodic dynamical system is only slightly changed, even in the infinite time limit. Consequences of this statistical analogue of the structural stability are: extension of the range of  practical applicability of the Boltzmann Ergodic Hypothesis, justification of the perturbation method in statistical physics, justification of the numerical approximations in molecular dynamic calculations and smoothness of the transition from bounded to unbounded motion as observed in numerical simulation of anomalous transport in tokamaks. 

\end{abstract}

\maketitle

%\baselineskip=22pt

%\date{29.04.2003}

%\preprint{PRL}


The stability of the predictions from equilibrium statistical physics is closely related on one hand to the Botzmann Ergodic Hypothesis (BEH) \cite{ErgThery} \cite{RBeq} and on the other hand to the persistence of the ergodic properties when a dynamical system (DS) is exposed to small perturbations. The accompanying continuity problem is related to the accuracy of the BEH in the neighborhood of a DS with proven ergodicity \cite{ChernSin} \cite{SimanyiNonlin}. 

There are known examples of DS's, for which the rational/irrational character of the parameters leads to a change in the behaviour from non chaotic to mixing \cite{GenTriangBillrds}. Weakly chaotic DS  \cite{zaslavski}, or DS at the verge of chaoticity \cite{JacquesRB}, are important for the study of anomalous transport in tokamaks. A basic problem related to perturbations is to guarantee the validity of the numerical approximations used to study the DS or, more generally, to prove the continuity of the infinite time limit with experimentally uncertain parameters. Such problem arises for example in area preserving maps ({\it e.g.} for the standard map \cite{Chirikov}) where the area of the phase space visited by the trajectories starting from a fixed small domain is a discontinuous function of the parameters. This behavior is the consequence of the destruction of invariant circles at well defined critical values of the stochasticity parameter.

We shall prove, using a main theorem, that a slightly perturbed DS defined in a finite phase space volume, despite not necessarily being ergodic, behaves almost like an ergodic DS when probed with observable relevant from the point of view of statistical physics. For the DS defined in a infinite phase space volume as in the case of the \emph{dissipative} DS \cite{Aaronson} ({\it i. e.} when the trajectories are unbounded with 100\% probability as for example in the case of the infinite ergodic DS's), the theorem justifies the absence of threshold in the transition from bounded to unbounded motion, as observed in numerical study of anomalous particle transport in tokamak \cite{JacquesRB}.

The continuity of the \emph{infinite time limit of the double, time and ensemble averages} (ITEA) is studied. It will be proved that this double average is stable under small perturbations in the ergodic and the dissipative cases. For DS that are not ergodic or dissipative, it is possible to have discontinuous behaviour, as in the previous example of the standard map. As a consequence of our analysis, unphysical number theoretic aspects of ergodic properties of weakly stochastic systems are clarified. The continuity results proved below is an additional argument in support to perturbation method in equilibrium statistical physics \cite{RBeq} and prove the stability of the results of microcanonical molecular dynamics simulations when the unperturbed system is ergodic.

%%%%

For the sake of simplicity and generality, we consider an abstract, discrete time, measure preserving, dynamical system \cite{ErgThery}. Such system may be defined on the phase space $M$ with the generalized phase space volume $\mu$ as conserved measure, and a discrete time \emph{one to one} evolution given by a (stroboscopic) map $\mathbf{x} \rightarrow T(\mathbf{x})$. Typically, $\mathbf{x} = \mathbf{(p,q)}$, and $d\mu(\mathbf{x}) = d\mathbf{p}d\mathbf{q}$. For the sake of mathematical rigor, we denote by $\mathcal{A}$ the family of the subsets of $M$, where $\mu$ is defined. For every subset $A$ of the family $\mathcal{A}$, we postulate that $\mu(A) = \mu \left[T(A)\right]$ \cite{ErgThery}, which is nothing else than the well known Liouville theorem in the case of Hamiltonian systems. We shall denote the DS by $(M,\mathcal{A},\mu,T)$. The measure can be either finite when it is normalized to $\mu(M)=1$, as in the case of the classical finite spatial extent DS with a potential energy bounded from below \cite{ErgThery}, or infinite like in the case of a two component classical plasma \cite{Aaronson}. The definition of an ergodic DS in the {\it mathematical literature}, for both the finite and the infinite cases, is the following: if a subset $A$ of $M$ is invariant, then either $A$ or its complement has zero measure. From this definition follows the equality of time and ensemble averages  \cite{ErgThery}, \cite{Aaronson}.

For any subset $A$ of $\mathcal{A}$, we denote by $v(T,A)$ the subset of $M$ visited by the trajectories of the map $T$, starting from $A$. The continuity of $\mu\lbrack\nu(T,A)]$ as a consequence of the main theorem will now be considered.

Without loss of generality, the relevant properties of the DS's from the statistical physics point of view can be studied, in a technically convenient manner, in Hilbert space $\mathcal{H}$ of square integrable functions $L^{2}(M,\mu)$ \cite{ErgThery} \cite{RBeq} with the scalar product defined as
$\left\langle \phi \, | \, \psi \right\rangle = \int_{M} \phi^{\ast}(\mathbf{x}) \, \psi(\mathbf{x}) \, d\mu(\mathbf{x})$. The linear evolution operator $U$ (also denoted $U(T)$) acting in $\mathcal{H}$ is defined by $\psi(\mathbf{x}) \rightarrow \psi\left[T(\mathbf{x)}\right] = (U\,\psi)(\mathbf{x})$. It is unitary due to the conservation of the measure $\mu$. Choosing $\phi(\mathbf{x})$ as the initial time probability density function (PDF) $\rho_{0}(\mathbf{x)}$, and $\psi(\mathbf{x})=Y(\mathbf{x})$ an observable, then $\langle\phi \, | \, U^{t} \, \psi\rangle = \langle\rho_{0} \, | \, U^{t} \, Y \rangle$ is the ensemble average of the observable $Y(\mathbf{x})$ at time $t$. Consequently, the physical meaning of the mathematical concept of weak convergence of DS can be made obvious. According to Refs. \cite{ErgThery} \cite{Aaronson}, the parametrized family of DS's $(M,\mathcal{A},\mu,T_{\varepsilon})$ converges weakly to the DS $(M,\mathcal{A},\mu,T)$ iff for any $\phi(\mathbf{x}),\psi(\mathbf{x})$ belonging to $\mathcal{H}$ it follows that  $\langle\phi \, | \, U(T_{\varepsilon}) \, \psi \rangle \mathbf{\rightarrow} \langle \phi \, | \, U(T) \, \psi\rangle$, when $\varepsilon$ $\rightarrow 0$. This weak convergence is denoted  $T_{\varepsilon}\overset{w}{\rightarrow}T$. We note that the weak convergence follows from the usual convergence of the maps $\mathbf{x} \rightarrow T_{\varepsilon}(\mathbf{x})$, but the converse, in general, is not true.

The elements of the one parameter family of DS's $(M,\mathcal{A},\mu,T_{\varepsilon})$ can be considered as weak perturbations of the DS $(M,\mathcal{A},\mu,T)$. This is the case, for instance, for hard ball systems obtained by adding a small potential energy or by inserting small reflecting obstacles of zero volume. It is easy to prove, using simple Hilbert space algebra, that $T_{\varepsilon }^{(k)} \overset{w} {\rightarrow} T^{(k)}$ is a consequence of $T_{\varepsilon} \overset{w} {\rightarrow} T$, where $T^{(k)}$ is the $k-th$ iterate of the map $T$. According to the definition and the physical interpretation of the weak convergence, it follows that for any abstract dynamical systems the \emph{time evolution of the ensemble average at any finite time has a continuous dependence on the weak perturbation of the unit time evolution}. This result being established, it remains to study the effect of a weak perturbation on the statistical properties of the infinite time limit.

The infinite time limit ($N\rightarrow\infty$) of the ensemble average $\langle \rho_{0} \, | \, U^{N} \, Y\rangle = \int_{M} \rho_{0}(\mathbf{x}) \, Y\left[T^{(N)}(\mathbf{x)}\right]  \, d\mu(\mathbf{x})$ in general does not exist at the exception of \emph{mixing} dynamical systems \cite{ErgThery}. Consequently we must restrict ourselves to the weaker information given by the ITEA, which is typically of the form 
\begin{eqnarray}
\lim_{N\rightarrow\infty}\frac{1}{N} \sum\nolimits_{n=0}^{N-1} \int_{M} \phi^{\ast}(\mathbf{x}) \, \psi\left[T^{(n)}(\mathbf{x)}\right] \, d\mu(\mathbf{x}) 
&&
\nonumber \\
= \lim_{N\rightarrow\infty } \frac{1}{N}\sum\nolimits_{n=0}^{N-1}\langle\phi|[U(T)]^{n}\psi\rangle && \, .
\nonumber
\end{eqnarray}
Recalling the von Neumann ergodic theorem this limit exists if $\phi(\mathbf{x})$ and $\psi(\mathbf{x})$ are square integrable \cite{ErgThery} \cite{Aaronson}.

Moreover, denoting by $\mathcal{H}_{inv}$ the subspace of $\mathcal{H}$ consisting of invariant functions, we have that $\psi(\mathbf{x)}$ belongs to $\mathcal{H}_{inv}$ iff $\psi(\mathbf{x}) = \psi\left[ T(\mathbf{x}) \right]$ ({\i.e.} iff $\psi=U(T)\,\psi$). In a similar way, denoting by $\mathcal{H}_{c}$ the orthogonal complement of $\mathcal{H}_{inv}$ {\it i.e.} $\mathcal{H}=\mathcal{H}_{inv} \oplus \mathcal{H}_{c}$ and defining the operator $P(T)$ as the orthogonal projector on the subspace $\mathcal{H}_{inv}$, the von Neumann ergodic theorem states:
\begin{equation}
\lim_{N\rightarrow\infty}\frac{1}{N}\sum\nolimits_{n=0}^{N-1}\langle
\phi|[U(T)]^{n}\psi\rangle=\langle\phi|P(T)\psi\rangle\label{vNeumann}
\end{equation}
In particular, if the initial time PDF is $\rho_{0}(\mathbf{x})$ then the function $\rho_{\infty}(\mathbf{x}) = [P(T)\rho_{0}](\mathbf{x})$ is the PDF of the points when the probability of a domain is calculated according to the mean visiting time fraction. The problem of the continuous dependence of the ITEA, in the sense of the weak convergence, reduces to the study of the following question: what are the conditions under which $T_{\varepsilon}\overset{w}{\rightarrow}T$ when $\varepsilon \rightarrow 0$ leads to $\langle\phi\,|\,P(T_{\epsilon}) \, \psi \rangle \rightarrow \langle \phi \, | \, P(T)\,\psi\rangle$ for all $\phi(\mathbf{x}),\psi(\mathbf{x})$ belonging to $\mathcal{H}$, {\it i.e.} $P(T_\epsilon) \overset{w}{\rightarrow} P(T)$.

A theorem is the starting point for a unified study of the finite or infinite DS. 

\vskip0.25truecm

\textit{Theorem 1: Consider the sequence of DS's $(M,\mathcal{A},\mu,T_{\varepsilon})$ that are weak perturbations of $(M,\mathcal{A},\mu,T)$, {\it i.e.} suppose that $T_{\varepsilon}\overset{w}{\rightarrow}T$. Then for all $\varphi(\mathbf{x})$ belonging to $\mathcal{H}$ and $\psi_{c}(\mathbf{x})$ belonging to $\mathcal{H}_{c}$, we have $\left\langle \varphi\,|\,[P(T_{\varepsilon})-P(T)]\,\psi
_{c}\right\rangle \rightarrow0$.} 

\vskip0.25truecm

The proof is performed in three steps: 

A) The statement $T_{\varepsilon}\overset{w}{\rightarrow}T$ is equivalent to $\left\langle \varphi_{1}\,|\,[U_{\varepsilon}-U]\,\varphi_{2}\right\rangle \rightarrow0$
\emph{for\ any} $\varphi_{1},\varphi_{2}$ belonging to $\mathcal{H}$. (We use here the shorter notations $U_{\varepsilon}=U(T_{\varepsilon})$, $U(T)=U$, $P(T_{\varepsilon})=P_{\varepsilon}$ and $P(T)=P$). Since $U_{\varepsilon}$ and $U$ are unitary, this statement is also equivalent to an apparently stronger statement: for \emph{any} $\psi(\mathbf{x})$ belonging to $\mathcal{H}$ we have $\left\Vert (U_{\varepsilon}-U)\,\psi\right\Vert \rightarrow0$ for $\varepsilon \rightarrow 0$. 

B) We denote $\eta(\varepsilon) = \left\vert \left\langle \varphi|(P_{\varepsilon}-P)\,\psi_{c}\right\rangle \right\vert$, then for $\psi_{c}\in\mathcal{H}_{c}$ {\it i.e.} $P \, \psi_{c}=0$ we have  $\eta(\varepsilon) =  \left\vert \left\langle \varphi|P_{\varepsilon}\,\psi_{c}\right\rangle \right\vert $. We now consider the normalization $\left\Vert \varphi\right\Vert =1$. \emph{The key lemma used to prove the von Neumann ergodic theorem \cite{ErgThery}} states that for $\psi_{c}$ \emph{belonging to} $\mathcal{H}_{c}
$ \emph{there exist a} $\chi_{\delta} \in \mathcal{H}$ \emph{such that} $\left\Vert \psi_{c}-(U-1)\emph{\ }\chi_{\delta}\,\right\Vert \leq\delta/2$. We obtain 
\begin{eqnarray}
\eta(\varepsilon) &\leq& \left\vert \langle
\varphi \, | \, P_{\varepsilon} \left[\psi_{c}-(U-1) \chi_{\delta}\,\right]  \rangle\right\vert 
\nonumber \\ 
&+& \left\vert \langle P_{\varepsilon} \varphi|\left[ (U-1)-(U_{\varepsilon}-1)\right]  \chi_\delta \rangle\right\vert
\nonumber \\ 
&+& \left\vert \langle\varphi\,|\,P_{\varepsilon}(U_{\varepsilon}-1)\,\chi_{\delta}\rangle\right\vert \, .
\nonumber
\end{eqnarray} 
Using the relation $P_\epsilon \, U_\epsilon = P_\epsilon$ \cite{ErgThery} the last term is zero. 

C) From Schwartz inequality it follows that 
$\eta(\varepsilon)\leq \delta/2+\left\Vert (U_{\varepsilon}-U)\chi_{\delta}\right\Vert \, .$ Indeed, from A) there exist a $\varepsilon$ such that $\left\Vert (U_{\varepsilon} - U)\chi_{\delta}\right\Vert \leq\delta/2$. Then, for $\delta$ arbitrary small
there exist a $\varepsilon$ such that $\left\vert \left\langle \varphi
|(P_{\varepsilon}-P)\,\psi_{c}\right\rangle \right\vert =\eta(\varepsilon
)\leq\delta$. This completes the proof.

We are now in position to prove a second Theorem valid for abstract ergodic systems. 

\vskip0.25truecm

\textit{Theorem 2: Consider the DS's $(M,{\mathcal{A}},\mu,T_{\varepsilon})$, $(M,{\mathcal{A}},\mu,T)$, $T_{\varepsilon}\overset {w}{\rightarrow}T$, and suppose that the DS $(M,{\mathcal{A}},\mu,T)$ is either ergodic (when $\mu(M)=1$), or dissipative (when $\mu(M)=\infty$). Then for all $\varphi(\mathbf{x})$ and
$\psi(\mathbf{x})$ belonging to $\mathcal{H}$ we have:}
\begin{equation}
\left\langle \varphi\,|\,[P(T_{\varepsilon})-P(T)]\,\psi\right\rangle
\rightarrow 0 \, {\rm for} \, \varepsilon \rightarrow 0
\label{weak_cont} \, .
\end{equation}

\vskip0.25truecm

Due to the linearity of the scalar product and the previously used Lemma it is sufficient to get a proof for the case when $\psi(\mathbf{x})=\psi_{inv}(\mathbf{x})$ is an invariant function. In our case $\psi_{inv}(\mathbf{x})\equiv const$ \cite{ErgThery}. We then have $U(T_{\epsilon}) \, \psi_{inv} = U(T) \, \psi_{inv} = \psi_{inv}$ and by using (\ref{vNeumann}) it follows that $P(T_{\epsilon}) \, \psi_{inv} = P(T) \, \psi_{inv} = \psi_{inv}$, which completes the proof.

According to the standard terminology of \emph{statistical physics}, a DS associated to an autonomous Hamiltonian systems with $N$ degree of freedom, is considered ergodic if the restriction of the motion to the hypersurfaces determined by a fixed energy is ergodic in \emph{mathematical} terms. We denote by $H(\mathbf{x})$ and $H_{\varepsilon}(\mathbf{x})$ the unperturbed (assumed ergodic) and the perturbed Hamiltonian functions, respectively. Then $\psi_{inv}(\mathbf{x})$ is of the form $\psi_{inv}(\mathbf{x}) = f(H(\mathbf{x}))$. Without loss of generality, we consider the smooth functions $\psi_{inv}(\mathbf{x})$ and denote the quantity $J_{\varepsilon}(\mathbf{x}) = f(H(\mathbf{x}))/f(H_{\varepsilon}(\mathbf{x}))$. Then, if in addition to the conditions of the Theorem $1$ we assume that $\left\vert J_{\varepsilon}(\mathbf{x})-1\right\vert <\delta(\varepsilon)/4$ with $\lim_{\varepsilon \rightarrow0}\delta(\varepsilon)=0$ (this happens, for instance, when $f(H(\mathbf{x}))=\exp(-\beta~H(\mathbf{x}))$), we obtain the physical result that: 

\vskip0.25truecm

\textit{An ergodic (in physical terms) Hamiltonian system satisfying the above conditions on $J_{\varepsilon}(\mathbf{x})$ obeys (\ref{weak_cont}).} 

\vskip0.25truecm

Indeed, recalling Theorem $1$, we only need to compute $P(T_{\varepsilon})\,\psi_{inv}$ and by the definition of $P(T_{\varepsilon}) = P_{\varepsilon}$ given by \ref{vNeumann}, we must estimate $(U_{\varepsilon }^{k}\psi_{inv})(\mathbf{x})$. Simple algebra then leads to $(U_{\varepsilon}^{k}\psi_{inv})(\mathbf{x}) = \psi_{inv}(\mathbf{x}) \, J_{\varepsilon}(T_{\varepsilon}^{(k)}(\mathbf{x)})/J_{\varepsilon}(\mathbf{x}) = \psi_{inv}(\mathbf{x}) \, L_{\varepsilon,k}(\mathbf{x})$. We also have $\left\vert L_{\varepsilon,k}(\mathbf{x}) - 1\right\vert <\delta(\varepsilon)$, for small $\delta(\varepsilon)$. From the normalization $\left\Vert \varphi\right\Vert = \left\Vert \psi_{inv}\right\Vert =1$ it follows that $\left\vert \left\langle \varphi\,|\,[U_{\varepsilon}^{k}-U^{k}]\,\psi_{inv}\right\rangle \right\vert <\delta(\varepsilon)$ and consequently that $\left\vert \left\langle \varphi \, | \, [P(T_{\varepsilon})-P(T)]\,\psi_{inv}\right\rangle \right\vert <\delta(\varepsilon)$. This result  completes the proof.

We shall denote in the following by $1_{A}(\mathbf{x})$ the \textit{function} equal to $1$ in the domain $A$ but zero outside (the characteristic function of $A$). If the system is ergodic, then $P(T) \, 1_{A}(\mathbf{x}) \equiv \mu(A) \, 1_{M}(\mathbf{x}) \equiv \mu(A)$. As an application, we consider $\mu(M)=1$ and $\varphi(\mathbf{x})=1_{S}(\mathbf{x})/\mu(S)$ the normalized distribution function of the starting points in the previous Eq. (\ref{weak_cont}). For simplicity, we consider the "source " of the particles localized in the domain $S$. We assume there is a "detector" in the domain $D$, measuring the mean visit time fraction, spent in $D$. According to \ref{vNeumann} this quantity is given, in general, by $\left\langle \varphi \ \,|\,P(T_{\varepsilon})\,1_{D}(\mathbf{x})\right\rangle $, which is close to
$\left\langle \varphi\ \,|\,P(T)\,1_{D}(\mathbf{x})\right\rangle $, or, due to the ergodicity of $T$, close to $\mu(D)$. Consequently, irrespectively of the positions of the "source" and of the "detector" and for sufficiently small perturbations the counting is almost the same as that performed on the unperturbed ergodic case. There are no possibility to protect the detector from the particles coming from the source. {\it If the system is not ergodic, then in general this property is no longer true}, as in the example of the near critical standard map. A striking example of non ergodic system having a sensible dependence on the perturbations is given in \cite{Vitot}. It is shown there how analytically calculated perturbations can lead to a drastic reduction of the particle transport in fusion plasma.

An alternative way to define ergodicity for finite measure DS ($\mu(M)=1$) is to impose $\mu(v(T,A))=1$ to every subset $A$ with $\mu(A)>0$. This means that the trajectories starting from an arbitrary small neighborhood of a phase space position completely fill the whole phase space. We prove the expected property: the trajectories of the perturbed DS will almost fill the entire phase space. To get the proof, we first observe that from Theorem $1$ we get a result on \emph{strong continuity}: 

\textit{Let $(M,\mathcal{A},\mu,T_{\varepsilon})$, $(M,\mathcal{A},\mu,T)$, $T_{\varepsilon}\overset{w}{\rightarrow}T$, and $(M,\mathcal{A},\mu,T)$ be ergodic, then for any $\psi(\mathbf{x})$ belonging to $\mathcal{H}$ we have $\left\Vert [P(T_{\varepsilon})-P(T)]\,\psi\right\Vert \rightarrow0$}. 

\vskip0.25truecm

This result is a direct consequence of (\ref{weak_cont}) and of the fact that $P(T_{\varepsilon})$ and $P(T)$ are projectors.

\vskip0.25truecm

From this remark, we get the following result on approximate ergodicity: 

\vskip0.25truecm

\textit{Let $(M,\mathcal{A},\mu,T_{\varepsilon})$, $(M,\mathcal{A},\mu,T)$, $T_{\varepsilon}\overset{w}{\rightarrow}T$, $\mu(M)=1$, and $(M,\mathcal{A}%
,\mu,T)$ ergodic. Consider a subset $S$ such that $\mu(S)>0$. Then $\mu(v(T_{\varepsilon},S))\rightarrow1$.} 

\vskip0.25truecm

This proof follows from the use the strong continuity, with $\psi(\mathbf{x})=1_{S}(\mathbf{x)}$, and from the fact that $P(T_{\varepsilon})1_{S}$ is zero outside the domain $v(T_{\varepsilon},S)$ and that $P(T)\,1_{S}(\mathbf{x)}$ is a nonzero constant function. This result shows that for a slightly perturbed ergodic DS the trajectories fill almost the same phase space as the trajectories would do  in the case of the unperturbed DS. Again, \emph{if the system is not ergodic then, in general, this property is no longer true.} A counterexample is again the standard map at the critical value of the stochasticity parameter.

For clarity, let the phase space $M$ be a torus, with coordinates $(p,q)$ defined modulo 1 and define a map $(p,q)\rightarrow T(p,q\mathbf{)}$ by $p \rightarrow p^{\prime}=p+\alpha$, $q\rightarrow q^{\prime}=q+\beta$ ($\operatorname{mod}~1$). When $\alpha$, $\beta$,
and $\alpha/\beta$ are all irrational, then the map is ergodic \cite{ErgThery}. Let the perturbed sequence $T_{\varepsilon}$ given by the non ergodic, rational approximations of $\alpha\approx\alpha_{\varepsilon}=N_{\varepsilon}/D_{\varepsilon}$, with irreducible fractions. If we choose two small circles of diameter $\delta$ for the source $S$ and the detector $D$, then if
$D_{\varepsilon}>1/\delta$, we get $v(T_{\varepsilon},A)=M$, so that the trajectories starting from $S$ visit every parts of the phase space. From this point of view, the system behaves like an ergodic one, but the frequency of visit of the detector approaches a limiting value only when $D_{\varepsilon}\gg1/\delta$. This example illustrates the fact that despite a weak perturbation of an ergodic DS does not necessarily lead to an ergodic DS, the latter behaves like an ergodic one.

\emph{The previous study of the very simple dynamics on the torus is of special interest because the dynamics associated to quasiperiodic time dependent Hamiltonian, with} $f$ \emph{ degrees of freedom, having} $N$ \emph{ frequencies, can be reformulated as an autonomous DS in an enlarged phase space, with} $2f+N$ \emph{dimensions, and} $N~$ \emph{\ ~cyclic coordinates:} $x=\{p_{1},...,p_{f},q_{1},...,q_{f},~\varphi_{1},...,\varphi_{N}\}$ \emph{and the} \textbf{invariant} \emph{measure} $d\mu(x)=d\mathbf{p}~d\mathbf{q}~d\mathbf{\varphi}$. \emph{This imbedding, known as skew product \cite{ErgThery}, allows for the use of the methods of ergodic theory.}

The deduction from Theorem $1$ is correct whenever the invariant functions of $T$ are constants and can be used also in the framework of \emph{non equilibrium statistical physics.} In the case of DS with $\mu(M)=\infty$, when $(M,\mathcal{A},\mu,T)$ is dissipative, the single invariant function is zero {\it i. e.} $P(T)\psi=0$. Then from $T_{\varepsilon}\overset{w}{\rightarrow}T$ and from \ref{weak_cont} it follows that $\left\langle \phi |\ P(T_{\varepsilon})\ \psi\right\rangle \longrightarrow 0$. As an example we  consider the transversal two dimensional guiding center motion of a charged particle in a constant magnetic field and in a fluctuating electric field reduced to the model proposed in \cite{JacquesRB}. The motion is described by a time periodic Hamiltonian system with a single degree of freedom. The typical phenomenon mentioned there is the transition form bounded motion to unbounded motion for large wave amplitudes with increasing wave amplitudes. 

Using numerical simulations, "the lack of stochasticity threshold and a smooth increase of the measure of the chaotic regions" (with increasing wave amplitudes) was discovered. To explain this effect, we assume that the trajectories evolving under the action of $T$ or of $T_{\varepsilon}$ start from the bounded domain $A$. We also assume that $T$ is dissipative (unbounded trajectories) and \textit{$T_{\varepsilon} \overset{w} {\rightarrow} T$}. In this case the invariant functions are zero and  Theorem $2$ holds. We have $P(T) 1_{A} \equiv 0$, and we assume {\it ad absurdum}, that the trajectories, under the action of $T_{\varepsilon}$ are all contained in the same bounded domain $B$. Then $\langle1_{M}|P(T_{\varepsilon})1_{A}\rangle=\langle P(T_{\varepsilon})1_{M}|1_{A}\rangle=\langle\ 1_{M}|1_{A}\rangle=\mu(A)~$. But $P(T_{\varepsilon})1_{A}=0$ outside of $B$, so $\langle1_{M} |P(T_{\varepsilon})1_{A}\rangle=\langle1_{B}|P(T_{\varepsilon})1_{A}\rangle$.
It follows that $\langle1_{B}|P(T_{\varepsilon})1_{A}\rangle=\mu(A)>0$. Recalling the continuity theorem we get $\left\langle 1_{B}|\left[  P(T_{\varepsilon})-P(T)\right] \ 1_{A}\right\rangle =\left\langle 1_{B}|\ P(T_{\varepsilon})\ 1_{A} \right\rangle \longrightarrow 0$. From this contradiction it follows that the transition to unbounded motion is not caused by a sudden opening of some transport barriers that protect the domain $A$. Since the quasiperiodic perturbations can be treated within our formalism, it follows that the same smooth transition will be observed in the case of electric perturbations with several incommensurate frequencies .

As a consequence of theorem $1$, it is proved, for ergodic and dissipative dynamic systems  exposed to small perturbation, that the properties that are relevant in the framework of the equilibrium and non equilibrium statistical physics are only weakly perturbed. In particular we
solved the problem of the dependence of the ergodic properties on number theoretic properties of the parameters. This continuity explains some aspects of the empirically observed behaviour of the ergodic and dissipative DS when exposed to small perturbations. If explored with coarse grained
instruments, the behavior of the perturbed ergodic system is almost ergodic. Together with the results from \cite{Vitot}, we conjecture that between the two extreme cases: from the ergodic to the complete integrable DS, there is an increased possibility to reduce the chaoticity. All of the results are consequences of the weakly continuous dependence on the perturbations of the projector on the invariant states.

%\section{Acknowledgments}

Discussions with R. Balescu, D. Carati (ULB, Brussels), J. H. Misguich
and J. D. Reuss (DRFC, CEA-Cadarache), M. Vittot (CNRS, Luminy) are greatly acknowledged.

\end{document}